\documentclass[conference]{IEEEtran}
\IEEEoverridecommandlockouts  
\usepackage{cite}

\usepackage{placeins}

\usepackage{soul}

\usepackage{array}
\usepackage{tabularx}

\usepackage{float}
\usepackage{subfig}

\usepackage{url}

\ifCLASSINFOpdf
\else
\fi
\usepackage{amsmath}
\usepackage{graphicx}
\usepackage{subcaption}

\usepackage{url}
\usepackage{graphicx} 

\usepackage{amssymb}

\usepackage{float} 
\usepackage{titlesec} 
\usepackage{xcolor}

\begin{document}
%
\title{Adaptive 5G Resource Allocation for Multistatic ISAC-Based UAV Detection and Tracking%
}

\author{%
    \IEEEauthorblockN{Cole Dickerson, Wahab Khawaja, Ismail G\"{u}ven\c{c}}
    \IEEEauthorblockA{Department of Electrical and Computer Engineering,\\
    North Carolina State University, Raleigh, NC, USA}
    \IEEEauthorblockA{Email: \{jcdicker, wkhawaj, iguvenc\}@ncsu.edu}
}


%


\maketitle
\begin{abstract}
Unmanned aerial vehicles (UAVs) enable numerous commercial and public-safety applications, yet they also create security risks near critical infrastructure, transportation hubs, and restricted airspace. While integrated sensing and communications (ISAC) can leverage existing wireless networks for UAV surveillance, practical deployment must address competition between sensing and communication demands, as well as the challenges associated with tracking highly maneuverable UAVs with low radar cross section (RCS). This paper investigates adaptive multistatic ISAC for load-aware UAV detection and tracking in 5G wireless networks. A shared-resource framework is developed to quantify how sensing waveform length, sensing transmission rate, and beam allocation affect communication throughput in a 5G new radio (NR) system. Detection performance is analyzed using Zadoff-Chu (ZC) sensing waveforms, while tracking continuity is evaluated through an $M$-of-$N$ detection model. To improve robustness under congestion, software-defined sensor (SDS) nodes exploit external signals of opportunity (SoO) to provide supplemental passive sensing opportunities when network resources become limited. Results show that adaptive sensing policies outperform fixed sensing reservations by preserving throughput under dynamic load while maintaining useful sensing capability. Under heavy congestion, SDS assistance substantially reduces tracking outage in the simulated scenarios. Cram\'er-Rao lower bound (CRLB) analysis demonstrates that multistatic sensing geometries improve localization accuracy and provide more uniform spatial coverage than monostatic sensing alone. These results highlight coordinated adaptive sensing and distributed multistatic support as a practical path toward resilient UAV surveillance in future wireless networks.
\end{abstract}


%
\IEEEpeerreviewmaketitle

\section{Introduction}

The rapid democratization of low-cost, high-capability small unmanned aerial vehicles (UAVs) has outpaced the development of ubiquitous surveillance infrastructure. Small drones are now widely used for delivery, inspection, agriculture, emergency response, and imaging, while advanced air mobility and autonomous aviation are expected to further increase low-altitude traffic density. Meanwhile, growing reports of unauthorized or unsafe UAV activity near airports, critical infrastructure, and other sensitive areas have highlighted the need for persistent detection, tracking, localization, and classification capabilities beyond traditional air traffic monitoring. As our airspace becomes more congested, avionics systems and unmanned traffic management (UTM) frameworks must safely coordinate both cooperative and non-cooperative aircraft \cite{faa2020utm}.

\subsection{Literature Review}

Existing UAV sensing solutions remain constrained by the limitations of individual sensing modalities \cite{azari2018key, 8337900}. Radar systems provide range and velocity information, but their performance can degrade in cluttered environments or under line-of-sight blockage, while dense deployments may become cost prohibitive \cite{10938573}. Radio frequency (RF)-based systems can detect signals radiated directly by drones, yet depend on identifiable transmissions and may be ineffective against autonomous or radio-silent platforms. Camera and electro-optical/infrared sensors enable visual identification, but are sensitive to lighting, weather, and occlusion, while acoustic sensors often suffer from limited range and environmental noise. Although multi-sensor fusion can significantly improve robustness, such systems often require dedicated infrastructure, careful calibration, and increased deployment complexity. Prior work has shown that fusing radar measurements with passive RF sensing can improve localization accuracy and tracking coverage relative to single-modality systems \cite{dickerson2025fusion}, while publicly available multimodal UAV datasets support the development and evaluation of data-driven sensing frameworks \cite{AADMdatasets}. Together, these observations motivate the use of network-assisted integrated sensing and communications (ISAC) approaches that leverage existing wireless infrastructure for scalable aerial surveillance.

5G cellular networks provide a compelling foundation for ISAC as a large-scale wireless infrastructure capable of wide-area sensing and communication without the cost of dedicated sensing deployments. Low-altitude UAV monitoring is emerging as a key 6G use case, further motivating cellular ISAC frameworks for UAV tracking while maintaining communication services \cite{11098638}. However, practical deployments still face several key challenges \cite{10726912, 10418473, 11021487, 10101848}. Sensing and communications must compete for finite radio resources, making static sensing allocations inefficient under time-varying load and limiting sensing opportunities during periods of congestion. In addition, some frameworks rely on custom radar-like waveforms that are difficult to integrate with standardized 5G operation. Networked ISAC architectures that incorporate adaptive coordination and unified resource allocation between sensing nodes remain an open area of research, alongside related considerations involving security and network resilience. These gaps motivate load-aware, standards-aligned, and multistatic ISAC strategies that preserve sensing performance under heavy traffic.

Recent work has explored 5G network-based passive radar for UAV detection using commercial cellular transmissions as illuminators of opportunity, where sensing performance depends on opportunistic communication occupancy rather than adaptive control of waveform and sensing resources \cite{10172437, 11443758}. Other studies have shown that existing 5G transmissions can be repurposed for passive sensing by estimating channel responses from standard reference signals, enabling low-cost distributed receivers that detect moving targets without consuming additional communication resources \cite{10908926}. In addition, 5G NR synchronization signal blocks (SSBs) have been investigated as structured and always-available illuminators for passive bistatic detection and localization, demonstrating that standardized cellular signaling can support sensing even when downlink traffic is sparse or absent \cite{5GNRSSB, 10083170}. While effective, these approaches offer limited control over sensing update rate, waveform structure, or resource availability. 

\begin{figure*}[!t]
    \centering
    \includegraphics[width=\textwidth]{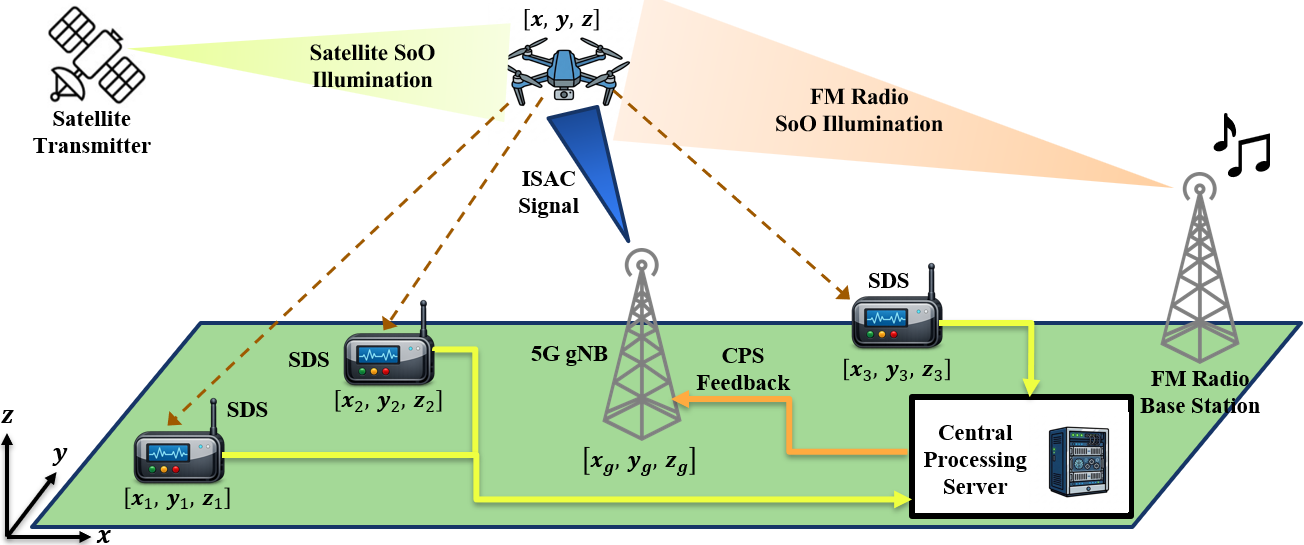}
    \caption{Proposed multistatic ISAC architecture for UAV detection and tracking. }
    \label{fig:NetwArch}
\end{figure*}

Cooperative 5G ISAC UAV sensing has shown that standardized cellular signals can support distributed UAV detection, localization, and tracking through coordinated network processing. Prior work proposed a two-stage 5G NR framework that uses beam-swept synchronization signals for coarse UAV detection and pilot-assisted links for refined tracking with centralized fusion, demonstrating standards-compatible network-assisted surveillance \cite{khawaja2026dyspan}. Other studies have leveraged coordinated beam sweeping, multistatic multiple-input multiple-output (MIMO) sensing, and direct downlink sensing with standard 5G NR signals to jointly estimate target parameters while preserving communication quality \cite{11443837, 10833700}. Complementary resource-allocation studies have further highlighted the importance of jointly optimizing waveform, beam, and radio resources under communication quality of service (QoS) constraints, while distributed cell-free and multistatic architectures improve sensing coverage, update freshness, and localization robustness \cite{11443754, 9921271, 10770016, 9945983}. However, frameworks that jointly address UAV tracking, adaptive sensing allocation, and distributed external support under realistic network congestion remain limited.

\subsection{Contributions}

To address these challenges, we propose an adaptive resource-allocation framework for multistatic ISAC that dynamically scales 5G sensing resources through configurable Zadoff-Chu (ZC) waveform and transmission parameters while leveraging software-defined sensors (SDSs) to exploit external signals of opportunity (SoO) for distributed sensing when network resources are constrained. The proposed approach builds on standardized 5G waveforms while extending sensing coverage, improving localization geometry, and maintaining sensing performance under heavy traffic. More specifically, our contributions are:
\begin{enumerate}
    \item We develop a load-aware resource-allocation model for 5G ISAC that quantifies sensing overhead as a function of ZC waveform length, sensing update rate, and beam count under shared radio resources;
    
    \item We characterize detection, throughput, and tracking tradeoffs under adaptive waveform scaling, demonstrating that 5G sensing preserves tracking capability while degrading gracefully as communication load increases;
    
    \item We introduce an SoO-assisted SDS multistatic sensing architecture that exploits external signals of opportunity, substantially reducing tracking outage when 5G sensing resources become scarce under congestion;
    
    \item We develop a Cram\'er-Rao lower bound (CRLB) framework for monostatic and multistatic ISAC sensing that captures waveform length, bandwidth, and coherent processing gain, demonstrating the localization and coverage benefits of distributed sensing.
\end{enumerate}

\section{Proposed ISAC System Architecture}

The proposed framework operates over the 5G NR air interface, where the serving gNB jointly supports downlink communication for user equipments (UEs) and sensing of targets such as UAVs. Specifically, the gNB schedules communications resources and transmits sensing waveforms, while distributed SDS nodes receive and process echoes reflected from the UAV.

Fig.~\ref{fig:NetwArch} illustrates the proposed multistatic ISAC architecture. The blue beam denotes an active sensing transmission from the 5G gNB, yellow and orange beams represent external signals of opportunity (SoO), and the brown dashed arrows indicate target reflections captured by SDS nodes. A target UAV is illuminated by multiple RF sources, including 5G gNBs and external SoO, such as satellite or FM broadcast transmissions. In addition to the serving gNB, the sensing infrastructure includes a distributed set of SDSs. These SDS nodes operate as passive receivers that capture target reflections without transmitting or participating in UE communications. While modeled here as passive radar receivers, SDS nodes could also function as passive RF sensors that detect and classify UAVs from telemetry or control-link emissions \cite{8913640}, and support localization through multi-node measurements using the sensing framework in \cite{ICCDickerson} and related datasets in \cite{raouf2025wirelessdatasets}. Based on network signaling, the SDS nodes can adapt their monitored frequency bands and sensing directions in real time. 

Measurements collected by the SDS network are forwarded to a central processing server (CPS), which fuses multi-source observations to generate a global situational awareness estimate. The resulting feedback is returned to the network to adapt sensing and communication parameters, including beamforming vectors, resource allocation, sensing bands, and transmit power. Updated coordination parameters may also be broadcast to SDS nodes through system information (SI), as discussed in Section~\ref{SDS}.

\subsection{ISAC Sensing Modalities: Active and Passive}

In active sensing mode, gNBs transmit 5G NR waveforms that jointly support user communications and target illumination for distributed SDS sensing. As controlled illuminators, the gNB transmissions can be coordinated by the CPS in terms of timing, waveform selection, and beamforming. Through the adaptive ZC sequence scaling described in Section~\ref{ZC_Adaption}, the network can increase processing gain and focus energy into narrow beams when the SDS network reports low signal-to-noise ratio (SNR), enabling more precise sensing and tracking. SDS nodes estimate target parameters via matched filtering and clutter suppression using the known transmitted waveform to mitigate direct-path interference, while the CPS fuses measurements across sensing modalities.

Coverage is further extended through passive sensing using external SoO. Unlike cellular waveforms, these sources are outside the control of the ISAC network, so their frequency, power, and modulation cannot be optimized for sensing. Nevertheless, their pervasive availability allows SDS nodes to exploit reflected SoO signals for persistent surveillance. This passive sensing layer can support target detection in areas with limited cellular coverage or during periods of heavy network congestion when 5G sensing resources are constrained.

\subsection{SDS Coordination and Sensing Operations}
\label{SDS}

The distributed SDS network relies on control signaling generated by the CPS and disseminated by the serving gNB to maintain situational awareness under varying network and RF conditions. This subsection describes the ISAC SI broadcast used to configure SDS nodes, their multi-band monitoring strategy, and directional sensing mechanisms.

\subsubsection{ISAC System Information}

\begin{figure}[!t]
    \centering
    \includegraphics[width=\columnwidth]{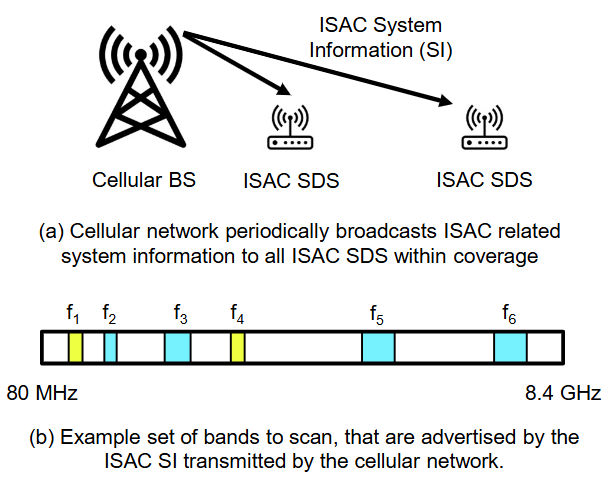}
    \caption{Illustration of ISAC SI signaling (a) and spectrum monitoring at an ISAC SDS (b).}
    \label{fig:ISAC_Bands}
\end{figure}

The control signaling mechanism for coordinating the distributed sensor array is illustrated in Fig.~\ref{fig:ISAC_Bands}(a). Based on CPS sensing-control decisions, 5G gNBs periodically broadcast ISAC SI to SDS nodes within coverage. This SI synchronizes passive SDS operation with active gNB transmissions and includes a priority field for predefined frequency bands, where zero disables monitoring and non-zero values determine the relative SDS duty cycle assigned to each band.

To reduce signaling overhead, SDS nodes exploit existing 5G procedures by extracting the Physical Cell Identify (PCI) from SSBs and retrieving static metadata, such as gNB locations and SoO waveform types, from local lookup tables. As a result, the proposed SI framework supports self-configuration of ISAC SDS nodes after deployment. The SI may be transmitted in multiple overhead tiers to balance network load and sensing precision. The low-overhead format provides binary monitoring status, the medium overhead format adds coarse UAV direction information for SDS receive steering, and the high-overhead format supports precision coherent sensing through fine angular coordinates and the precoding matrix index of the sensing waveform. Example SI formats are summarized in Table~\ref{tab:isac_si_formats}.

\begin{table}[t]
\centering
\caption{ISAC SI message formats by overhead levels.}
\label{tab:isac_si_formats}
\begin{tabular}{|p{35pt}|p{45pt}|c|p{100pt}|}
\hline
\textbf{SI Format} & \textbf{Field} & \textbf{Bits} & \textbf{Logic \& Description} \\ \hline
\textbf{Low Overhead} & Priority Level & 8 & 1-bit per band (0: Off, 1: On) \\ \cline{2-4} 
 & \textbf{Total} & \textbf{8} & Minimal overhead \\ \hline \hline
\textbf{Medium Overhead} & Priority Level & 16 & 2-bits per band (4 tiers of scheduling) \\ \cline{2-4} 
 & UAV Loc. (Coarse) & 12 & 6-bit Azimuth, 6-bit Elevation with respect to this BS \\ \cline{2-4} 
 & \textbf{Total} & \textbf{28} & Directional tracking with coarse spatial data \\ \hline \hline
\textbf{High Overhead} & Priority Level & 24 & 3-bits per band (8 tiers of scheduling) \\ \cline{2-4} 
 & UAV Loc. (Fine) & 24 & 12-bit azimuth, 12-bit elevation with respect to this BS (or coarse tracking of 2 UAVs)\\ \cline{2-4} 
 & Precoding Index & 12 & BS transmit Precoding Matrix Index (PMI) \\ \cline{2-4} 
 & \textbf{Total} & \textbf{60} & Precision coherent multistatic sensing \\ \hline
\end{tabular}
\end{table}

\subsubsection{Multi-Band Monitoring}

Fig.~\ref{fig:ISAC_Bands}(b) illustrates an example monitoring set containing six frequency bands. Yellow bands denote SoO, while blue bands denote cellular carriers such as 5G NR or 4G LTE. Because each SDS has limited instantaneous capture bandwidth, it cannot monitor all advertised bands simultaneously. Instead, the SDS sweeps across selected bands according to a duty cycle, dwelling on each band before switching to the next. 

This sequential monitoring creates a trade-off between spectral breadth and sensing latency. Wider capture bandwidth can reduce sweeping requirements at the cost of increased hardware complexity. Narrow bandwidth lowers cost, but increases revisit time as more bands are monitored, delaying detection or tracking updates within any individual band. Conversely, monitoring fewer bands improves temporal resolution but risks missing target reflections in unobserved spectrum. Therefore, the SDS must balance these constraints, potentially prioritizing selected bands using the ISAC SI indicators.

\subsection{Adaptive Sensing Waveform Design}
\label{ZC_Adaption}

The active sensing capability of the proposed framework is enabled through adaptive waveform design at the gNB. For a 5G ISAC system, the $u$-th root ZC sequence of length $N_{\rm ZC}$ is defined as:
\begin{equation}
x_u(n)=\exp\left(-j\frac{\pi u n (n+d_{\rm mod})}{N_{\rm ZC}}\right), \quad 0 \le n < N_{\rm ZC},
\end{equation}
where $u$ is the root index, $n$ is the sample index, and $d_{\rm mod}$ denotes the sequence parity. The sequence length $N_{\rm ZC}$ is typically chosen to be prime, with $u$ relatively prime to $N_{\rm ZC}$. In the proposed framework, the gNB adapts $N_\text{ZC}$ as one of several sensing-control variables, alongside sensing rate, beamforming, and transmit power. Since $N_\text{ZC}$ determines the processing gain (as described in Section~\ref{sensingmodel}), larger sequence lengths improve detection of smaller or more distant UAVs, but require more physical resource blocks (PRBs).

In addition to coherent processing gain, increasing $N_\text{ZC}$ also increases the occupied sensing bandwidth since additional PRBs and subcarriers are allocated to the sensing waveform. For a sensing waveform occupying $N_\text{PRB}$ PRBs, the corresponding bandwidth is $\beta = N_\text{PRB}N_\text{SC} \Delta f$, where $N_\text{SC}$ is the number of subcarriers per PRB and $\Delta f$ is the OFDM subcarrier spacing. Since radar range resolution improves with bandwidth, larger sensing waveforms also improve the approximate range resolution, $\Delta R \approx c/2\beta$, where $c$ is the speed of light. Therefore, increasing $N_\text{ZC}$ improves both detection sensitivity and range resolution at the cost of greater sensing-resource consumption.

In 5G NR, each PRB contains 12 subcarriers. Because ZC lengths are typically chosen as primes, they do not align exactly with multiples of 12, resulting in discrete PRB allocation steps and a small number of unused subcarriers. Table~\ref{tab:zc_scaling_overhead} summarizes representative sequence lengths, occupied bandwidths, range resolutions, and coherent processing gains.

\begin{table}[t] \centering \caption{$N_\text{ZC}$, sensing bandwidth, range resolution, and processing gain.} \label{tab:zc_scaling_overhead} \begin{tabular}{|c|c|c|c|c|} \hline \textbf{$N_{\rm ZC}$} & \textbf{PRBs} & \textbf{BW (MHz)} & \textbf{Range Res. (m)} & \textbf{Gain (dB)} \\ \hline 71 & 6 & 1.08 & 138.9 & 18.5 \\ \hline 139 & 12 & 2.16 & 69.4 & 21.4 \\ \hline 211 & 18 & 3.24 & 46.3 & 23.2 \\ \hline 283 & 24 & 4.32 & 34.7 & 24.5 \\ \hline 419 & 35 & 6.30 & 23.8 & 26.2 \\ \hline 503 & 42 & 7.56 & 19.8 & 27.0 \\ \hline 631 & 53 & 9.54 & 15.7 & 28.0 \\ \hline 839 & 70 & 12.60 & 11.9 & 29.2 \\ \hline 1063 & 89 & 16.02 & 9.36 & 30.3 \\ \hline 1291 & 108 & 19.44 & 7.72 & 31.1 \\ \hline \end{tabular} \end{table}

\subsubsection{PRB Scheduling Strategy}

The ZC sequence length is adaptively selected according to the observed sensing conditions and current network load while respecting available PRB constraints. The system monitors the observed SNR of backscattered echoes and the current network load $(\eta)$. After target detection, the CPS prioritizes tracking continuity by using the strongest sensing configuration permitted under the current load. Sensing resources are reduced only when higher-priority communication traffic limits available PRBs.

\begin{figure}[!t]
    \centering
    \includegraphics[width=\columnwidth]{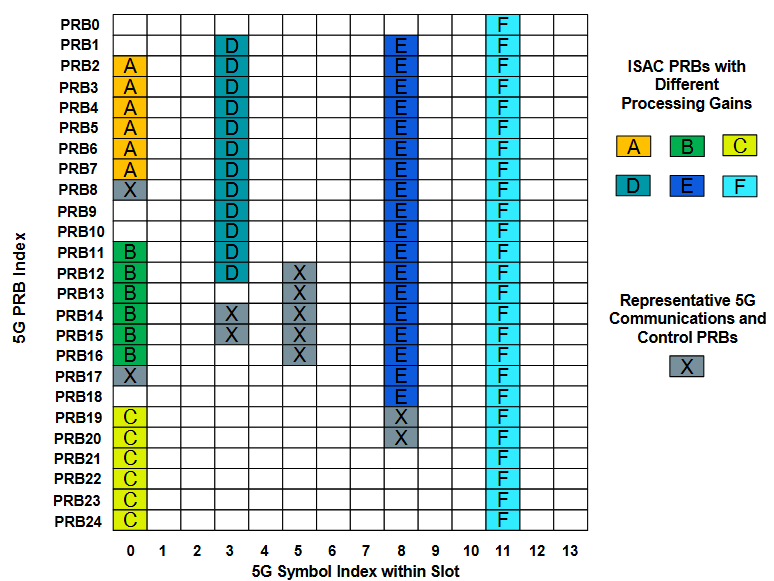}
    \caption{PRB allocation for sensing and communication.}
    \label{fig:PRBs}
\end{figure}

An example allocation for a 5G carrier with 5 MHz bandwidth and 15 kHz subcarrier spacing over one slot (1 ms) and 14 OFDM symbols is shown in Fig.~\ref{fig:PRBs}. In this example, the gNB schedules three ZC sequences of roots A, B, and C in the first symbol, each occupying 6 PRBs. Additional ZC sequences with roots D, E, and F, providing higher processing gain, are scheduled later in the slot to track detected UAVs. Representative PRB allocations for communication and control channels are denoted by PRB X. Under high network load, all subcarriers may be occupied by user data and control traffic.

\subsubsection{Directional Multi-Root Search and Precoding Strategy}

\begin{figure}[t!]
\centering
\includegraphics[width=0.99\columnwidth]{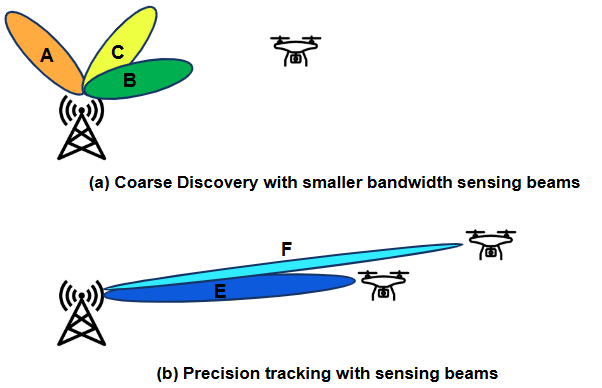}
\caption{Coarse search and precision tracking with ISAC beams.}
\label{fig:tracking}
\end{figure}

Examples of PRB precoding for coarse discovery (A,B,C) and precision tracking (E,F) are shown in Fig.~\ref{fig:tracking}. Similar to UE-specific beamforming in LTE and 5G, sensing PRBs may be precoded toward candidate sectors or an identified UAV direction. The proposed framework employs a hierarchical directional search with two operating modes: coarse discovery and precision tracking (as in \cite{khawaja2026dyspan}). During coarse discovery, the gNB transmits shorter ZC sequences (e.g., $N_\text{ZC} = 71 \text{ or }139$) using multiple root indices mapped to spatial precoders, allowing wide-area sector scanning with low resource overhead. 

When a candidate target is detected at angle $\hat{\theta}$, the system transitions to precision tracking. The gNB allocates a larger ZC sequence and applies a narrow high-gain beam directed toward $\hat{\theta}$. The longer sequence improves coherent processing gain, effective sensing SNR, and range resolution, while directional precoding improves received target power, angular estimation accuracy, and interference suppression. After target detection, the CPS may increase the sensing transmission rate (serving as an effective pulse-repetition-frequency) while maintaining the focused beam. The resulting higher slow-time sampling rate improves Doppler and micro-Doppler resolution for UAV classification  \cite{11443758, 11509266}. This two-stage strategy enables persistent wide-area surveillance while reserving high sensing resources for target confirmation and tracking.

\section{System Model and Performance Framework}

\subsection{Geometry Model and CRLB Analysis}
\label{geometry}

Consider a three-dimensional surveillance region containing one target UAV, one serving gNB, and a set of $N$ SDSs. The unknown UAV position is represented by the vector $\mathbf{x} =[x,\,y,\,z]^T$ and its estimated location is $\hat{\mathbf{x}} = [\hat{x},\hat{y},\hat{z}]^T$. The serving gNB is located at the known position $\mathbf{x}_g =[x_g,\, y_g,\,z_g]^T$, and the $i$-th SDS is deployed at the known location $\mathbf{x}_i = [x_i,\, y_i,\,z_i]^T$ for $i=1,...,N.$ The primary focus of this work is the resulting single-gNB multistatic sensing architecture. As a comparison baseline, we also model a conventional monostatic sensing configuration in which the serving gNB acts as both transmitter and receiver. Extension to cooperative multi-gNB deployments is left for future work.

Let $d_g \triangleq\left\|\mathbf{x}-\mathbf{x}_g\right\|_2$ and $d_i \triangleq\left\|\mathbf{x}-\mathbf{x}_i\right\|_2$ for $i=1,...,N,$ where $d_g$ is the one-way UAV-to-gNB range and $d_i$ is the UAV-to-$i$-th SDS range. For a receiving sensing node $s$ (either gNB or SDS) located at $\mathbf{x}_s = [x_s,\,y_s,\,z_s]^T$, define the coordinate differences as $\Delta x_s = x - x_s$, $\Delta y_s = y - y_s$, $\Delta z_s = z - z_s$. The corresponding horizontal (2D) and Euclidean (3D) distances are given by $\rho_s = \sqrt{\Delta x_s^2 + \Delta y_s^2}$ and $r_s = \sqrt{\Delta x_s^2 + \Delta y_s^2 + \Delta z_s^2}$, respectively. The azimuth and elevation angles to the UAV are given by $\theta=\operatorname{atan} 2\left(\Delta y_s, \Delta x_s\right)$ and $\phi=\operatorname{atan} 2\left(\Delta z_s, \rho_s\right)$. 

For the monostatic sensing mode, the observing node is the gNB ($\mathbf{x}_s = \mathbf{x}_g$) and the range $r = d_g$. For the $i$-th bistatic sensing link, the observing node is the $i$-th SDS ($\mathbf{x}_s = \mathbf{x}_i)$ and the bistatic range $r = d_g + d_i$. The generalized observation model is $\mathbf{z} = \mathbf{h}(\mathbf{x}) + \mathbf{n},$ where $\mathbf{z} = [z_r,\,z_\theta,\,z_\phi]^T$ is the observed measurement vector, $\mathbf{h}(\mathbf{x}) =  [r,\,\theta,\,\phi]^T$ is the noiseless measurement vector determined by the UAV position $\mathbf{x}$, and $\mathbf{n} = [n_r,\,n_\theta,\,n_\phi]^T$ is the measurement noise vector, where $n_{r},\,n_{\theta},\, \text{and }n_{\phi}$ denote the range, azimuth, and elevation noise terms, respectively. The measurement noise is modeled as zero-mean Gaussian with covariance $\mathbf{R}_i(N_\text{ZC})$, i.e., $\mathbf{n} \sim \mathcal{N}(\mathbf{0}, \mathbf{R}_i(N_\text{ZC}))$, where $\mathbf{R}_i(N_\text{ZC})$ depends on the sensing-link geometry, waveform parameters, and effective SNR.

For the monostatic sensing case, the received echo input SNR follows the conventional radar equation:
\begin{equation}
{\rm SNR_\text{in}}
=
\frac{
P_t G_t G_r \lambda^2 \sigma_{\text{RCS}}
}{
(4\pi)^3 d_g^4 kT \beta F L
},
\end{equation}
where $P_t$ is the transmit power, $G_t$ and $G_r$ are the transmit and receive antenna gains, $\lambda$ is the carrier wavelength, $\sigma_{\text{RCS}}$ is the UAV radar cross section, $k$ is Boltzmann's constant, $T$ is the system noise temperature, $F$ is the receiver noise figure, and $L$ captures additional system losses. For the bistatic sensing case, the monostatic propagation-loss term $d_g^4$ is replaced by the bistatic propagation-loss term
$d_g^2 d_i^2$. Incorporating the coherent processing gain associated with the selected ZC waveform length $N_\text{ZC}$, the effective sensing SNR is given by ${\rm SNR_{\text{eff}}} = \rm SNR_\text{in} \cdot N_\text{ZC}$ under noise-limited sensing conditions; interference-limited operation may yield smaller gains.

Following the standard CRLB for time-delay estimation in additive white Gaussian noise, the corresponding monostatic and bistatic range variances are given by:
\begin{equation}
\sigma_{r,m}^{2}
\approx
\frac{c^2}{32\pi^2\beta^2{\rm SNR}_{\rm eff}},
\quad
\sigma_{r,b,i}^{2}
\approx
\frac{c^2}{8\pi^2\beta^2{\rm SNR}_{\rm eff}}.
\end{equation}
The azimuth and elevation variances are similarly modeled as inversely proportional to the effective sensing SNR:
\begin{equation}
\sigma_{\theta,i}^{2}
=
\frac{\kappa_{\theta,i}}{{\rm SNR}_{\rm eff}},
\quad
\sigma_{\phi,i}^{2}
=
\frac{\kappa_{\phi,i}}{{\rm SNR}_{\rm eff}},
\end{equation}
where angular variance scaling parameters $\kappa_{\theta,i}$ and $\kappa_{\phi,i}$ capture array
aperture, beamwidth, and estimator-dependent effects. 

To quantify the best achievable UAV localization accuracy for the proposed sensing geometry, the CRLB is employed as a theoretical lower bound on the covariance of any unbiased estimator \cite{kay1993fundamentals}. In practice, range and angle estimates may be biased by hardware and signal processing imperfections; however, for tractability and to isolate sensor geometry effects, measurement errors are modeled as independent zero-mean Gaussian variables about the true range, azimuth, and elevation values. The Fisher Information Matrix (FIM) is given by:
\begin{equation}
    \mathbf{J}(\mathbf{x},N_\text{ZC})=\mathbf{H}^T(\mathbf{x}) \mathbf{R}_{i}^{-1}(N_\text{ZC}) \mathbf{H}(\mathbf{x}), 
\end{equation}
where $\mathbf{H}(\mathbf{x}) = \partial \mathbf{h}(\mathbf{x}) / \partial \mathbf{x}$ is the Jacobian matrix. For the monostatic case $(\mathbf{x}_s = \mathbf{x}_g)$, the Jacobian is:
{\small
\setlength{\arraycolsep}{2pt}
\begin{equation}
\mathbf{H}_m(\mathbf{x})=
\begin{bmatrix}
\tfrac{\partial r_g}{\partial x} & \tfrac{\partial r_g}{\partial y} & \tfrac{\partial r_g}{\partial z}\\
\tfrac{\partial \theta_g}{\partial x} & \tfrac{\partial \theta_g}{\partial y} & \tfrac{\partial \theta_g}{\partial z}\\
\tfrac{\partial \phi_g}{\partial x} & \tfrac{\partial \phi_g}{\partial y} & \tfrac{\partial \phi_g}{\partial z}
\end{bmatrix}
=
\begin{bmatrix}
\tfrac{\Delta x_g}{r_g} & \tfrac{\Delta y_g}{r_g} & \tfrac{\Delta z_g}{r_g}\\
-\tfrac{\Delta y_g}{\rho_g^2} & \tfrac{\Delta x_g}{\rho_g^2} & 0\\
-\tfrac{\Delta x_g\Delta z_g}{r_g^2\rho_g} &
-\tfrac{\Delta y_g\Delta z_g}{r_g^2\rho_g} &
\tfrac{\rho_g}{r_g^2}
\end{bmatrix}.
\end{equation}
}
For the $i$-th bistatic sensing link, only the first row of Jacobian $\mathbf{H}_{b,i}(\mathbf{x})$ differs from the monostatic case, reflecting the different range measurement function in the bistatic case:
{\small
\begin{equation}
\begin{aligned}
\mathbf{h}_{b,i}^{(1)}(\mathbf{x)}
&=
\left[
\tfrac{\partial r_{b,i}}{\partial x}\;\;
\tfrac{\partial r_{b,i}}{\partial y}\;\;
\tfrac{\partial r_{b,i}}{\partial z}
\right] \\[4pt]
&=
\left[
\tfrac{x-x_g}{r_g}+\tfrac{x-x_i}{r_i}\;\;
\tfrac{y-y_g}{r_g}+\tfrac{y-y_i}{r_i}\;\;
\tfrac{z-z_g}{r_g}+\tfrac{z-z_i}{r_i}
\right].
\end{aligned}
\end{equation}
}
Under the independent Gaussian measurement assumption, the per-node covariance matrix is  $\mathbf{R}_i(N_\text{ZC})=\operatorname{diag}\left(\sigma_{r, i}^2, \sigma_{\theta, i}^2, \sigma_{\phi, i}^2\right)$. Assuming independent sensing-node observations, the FIM for a given sensing configuration is:
\begin{equation}
\mathbf{J}(\mathbf{x},N_{\text{ZC}})
=
\sum_{i \in \mathcal{S}}
\mathbf{H}_{i}^{T}(\mathbf{x})
\mathbf{R}_{i}^{-1}(N_{\text{ZC}})
\mathbf{H}_{i}(\mathbf{x}),
\end{equation}
where $\mathcal{S}=\{\text{gNB}\}$ for the monostatic configuration and
$\mathcal{S}=\{1,\ldots,N\}$ for the multistatic SDS configuration. The CRLB for the UAV position estimate is $\mathbf{C}_{\mathrm{CRLB}}(\mathbf{x},N_\text{ZC})=\mathbf{J}^{-1}(\mathbf{x},N_\text{ZC})$. A scalar measure of localization accuracy is given by the position error bound (PEB), $\operatorname{PEB}(\mathbf{x}, N_\text{ZC})=\sqrt{\operatorname{tr}\left(\mathbf{J}^{-1}(\mathbf{x},N_\text{ZC})\right)}$, which is the lower bound on the achievable three-dimensional root-mean-square error (RMSE) for any unbiased estimator: $\operatorname{RMSE}(\mathbf{x},N_\text{ZC}) \geq \operatorname{PEB}(\mathbf{x},N_\text{ZC})$.

\subsection{Communication Resource Model}
\label{commsmodel}

While Section~\ref{geometry} characterizes sensing geometry and localization accuracy, we next model how shared communication resources constrain sensing opportunities. We consider a 5G NR system in which communication and sensing share a common time-frequency resource grid. Let $N_\text{{PRB,total}}$, $N_\text{{sym/sec}}$, $f_\text{{sense}}$, $N_b$, and $\mathrm{OH}$ denote the total number of available PRBs, OFDM symbols per second, sensing transmission rate, number of sensing beams, and fraction of resources reserved for mandatory NR signaling and control overhead, respectively. The remaining fraction, $1-\mathrm{OH}$, constitutes the shared resource budget over which communication traffic and active sensing compete. Since each PRB contains 12 subcarriers, the number of PRBs occupied by one sensing waveform is $N_\text{PRB,sense} = \left\lceil N_{\text{ZC}}/12 \right\rceil$. Under a worst-case linear beam-scaling assumption, where each active sensing beam occupies independent resources, the fraction of the post-overhead shared resource budget consumed by active sensing per second is given by: 
\begin{equation}
\alpha_{\text{sense}} =
\left(\frac{N_b}{1-\mathrm{OH}}\right)
\left(\frac{f_{\text{sense}}}{N_{\text{sym/sec}}}\right)
\left(\frac{N_{\text{PRB,sense}}}{N_{\text{PRB,total}}}\right).
\label{eq:multibeam}
\end{equation}
This provides a conservative upper bound on sensing overhead, since practical deployments may reduce resource usage through beam reuse, shared illumination, adaptive update rates, or multi-beam precoding. Multiple targets may also share a sensing beam provided they remain separable in the range-Doppler domain. Based on this framework, we consider four resource-allocation policies representing different operating philosophies under increasing communication load:
\begin{enumerate}
    \item \textit{Communications-Oriented Policy:} Maintains a fixed low sensing reservation to prioritize throughput at the expense of detection and tracking performance.
    \item \textit{Sensing-Oriented Policy:} Maintains a fixed high sensing reservation to prioritize sensing robustness at the expense of communication capacity.
    \item \textit{Adaptive 5G ISAC Policy:} Selects the largest feasible ZC sequence length under the current load while satisfying communication resource constraints.
    \item \textit{Adaptive 5G ISAC with SoO Assistance:} Uses SDS nodes to exploit external signals of opportunity and support tracking under heavy load while reducing dependence on active 5G sensing resources.
\end{enumerate}

Let $\eta \in [0, 1]$ denote the offered communication load normalized by the post-overhead shared resource budget $1-\mathrm{OH}$. Active 5G sensing is feasible only if:
\begin{equation}
    \alpha_\text{{sense}} \leq1-\eta.
\end{equation}
Let $N_{Z C}^{\star}(\eta)$ denote the adaptively selected ZC sequence length under load $\eta$. The adaptive policy selects:
\begin{equation}
    N_{\text{ZC}}^{\star}(\eta)=\max \left\{N_{\text{ZC}} : \alpha_{\text {sense }} \leq1-\eta\right\} .
\end{equation}
The normalized served throughput (assuming fixed spectral efficiency and bandwidth) is modeled as:
\begin{equation}
    T(\eta)=\min \left\{\eta,1-\alpha_{\text {sense }}\right\},
\end{equation}
where $\alpha_\text{sense}$ is expressed as a fraction of the post-overhead shared resource budget. For the adaptive policy, $N_{\text{ZC}}=N_{\text{ZC}}^{\star}(\eta)$. For fixed-budget policies, $N_{\text{ZC}}$ is held constant according to the selected sensing reservation. In the SoO-assisted case, sensing is progressively shifted from active 5G transmissions to passive SoO observations when constrained sensing resources prevent the tracker from satisfying the track-continuity requirements defined in Section~\ref{sensingmodel}, enabling the served throughput to match the offered load, i.e., $T(\eta)=\eta$.

\subsection{Detection and Tracking Model}
\label{sensingmodel}

Given the resource-allocation model in Section~\ref{commsmodel}, we next quantify how sensing waveform selection and sensing opportunities determine detection and tracking reliability. Using the effective sensing SNR model introduced in Section~\ref{geometry}, the probability of detection for a target false alarm probability $P_{\text{fa}}$ is modeled using the first-order Marcum $Q$-function:
\begin{equation}
P_{\text{d}}
=
Q_1\left(
\sqrt{2{\rm SNR}_{\rm eff}},
\sqrt{-2 \ln P_{\text{fa}}}
\right).
\end{equation}
The corresponding probability of missed detection is given by $P_{\text{md}} = 1 - P_{\text{d}}.$ 

Over a tracking observation window, let $N_{\text{5G}}$ denote the number of active 5G sensing opportunities and let $M$ denote the minimum number of successful detections required to maintain a track. Treating each sensing opportunity as an independent Bernoulli trial with success probability $P_{\text{d}}$, the number of successful active detections satisfies $K_{5 G} \sim \operatorname{Binomial}\left(N_{\text {5G }}, P_{\text{d}}\right)$. Accordingly, the tracking success probability for active 5G sensing is:
\begin{equation}
    P_{\mathrm{cont}}^{5 G}=\sum_{k=M}^{N_{\mathrm{5G}}}\binom{N_{\mathrm{5G}}}{k} P_d^k\left(1-P_d\right)^{N_{\mathrm{5G}}-k},
\end{equation}
and the corresponding outage probability is $P_{\mathrm{out}}^{5 G}=1-P_{\mathrm{cont}}^{5 G}$. Communication load affects tracking through the feasible sensing budget in Section~\ref{commsmodel}, which determines the selected waveform length $N_{\text{ZC}}^{\star}(\eta)$ and thus the resulting detection probability $P_{\text{d}}(N_{\text{ZC}}^{\star}(\eta)).$

For the SoO-assisted case, let $N_{\text{SoO}}$ denote the number of passive sensing opportunities within the same observation window and let $P_{\text{d,SoO}}$ denote the effective detection probability of each SoO opportunity. Then $K_{\text{SoO}} \sim \operatorname{Binomial}\left(N_{\text {SoO}}, P_{\text{d},\text{SoO}}\right)$. The total number of successful detections is modeled as the sum of two independent binomial random variables: $K_{\text{tot}} = K_{\text{5G}} + K_{\text{SoO}}$. Assuming independent active 5G and SoO sensing outcomes, the SoO-assisted tracking success probability becomes: $P_{\mathrm{cont}}^{\mathrm{SoO}}=\operatorname{Pr}\left(K_{\mathrm{tot}} \geq M\right)$, where the distribution of $K_{\text{tot}}$ is obtained by convolving the two binomial probability mass functions. The corresponding outage probability is $P_{\mathrm{out}}^{\mathrm{SoO}}=1-P_{\mathrm{cont}}^{\mathrm{SoO}}$. These models enable comparative evaluation of the proposed resource-allocation policies under varying communication load.

\section{Numerical Results}

We evaluate the proposed framework via simulations of sensing performance, communication-load tradeoffs, tracking reliability, and localization accuracy under representative sensing geometries. Simulations assume a 5G NR FR1 system at 3.5 GHz (band n78) with 50 MHz channel bandwidth, 15 kHz subcarrier spacing, normal cyclic prefix, and $N_\text{PRB,total}=270$ available PRBs per slot over the carrier bandwidth, consistent with 3GPP NR numerology \cite{3gpp38104}. Each PRB contains 12 subcarriers, and each 1 ms slot consists of 14 OFDM symbols. Candidate sensing waveform lengths and corresponding occupied sensing bandwidths follow Table~\ref{tab:zc_scaling_overhead}, while the overall carrier bandwidth remains fixed. The nominal NR overhead is $\mathrm{OH}=0.14$, following the FR1 downlink overhead model in \cite{3gpp38306}. Unless stated otherwise, $f_\text{sense} = 500$ Hz and the tracking model assumes an $M$-of-$N$ criterion with $M=3$ detections required over $N=5$ sensing opportunities to maintain track continuity.

\subsection{Sensing-Resource Overhead}

\begin{figure}[!t]
    \centering
    \includegraphics[width=\columnwidth]{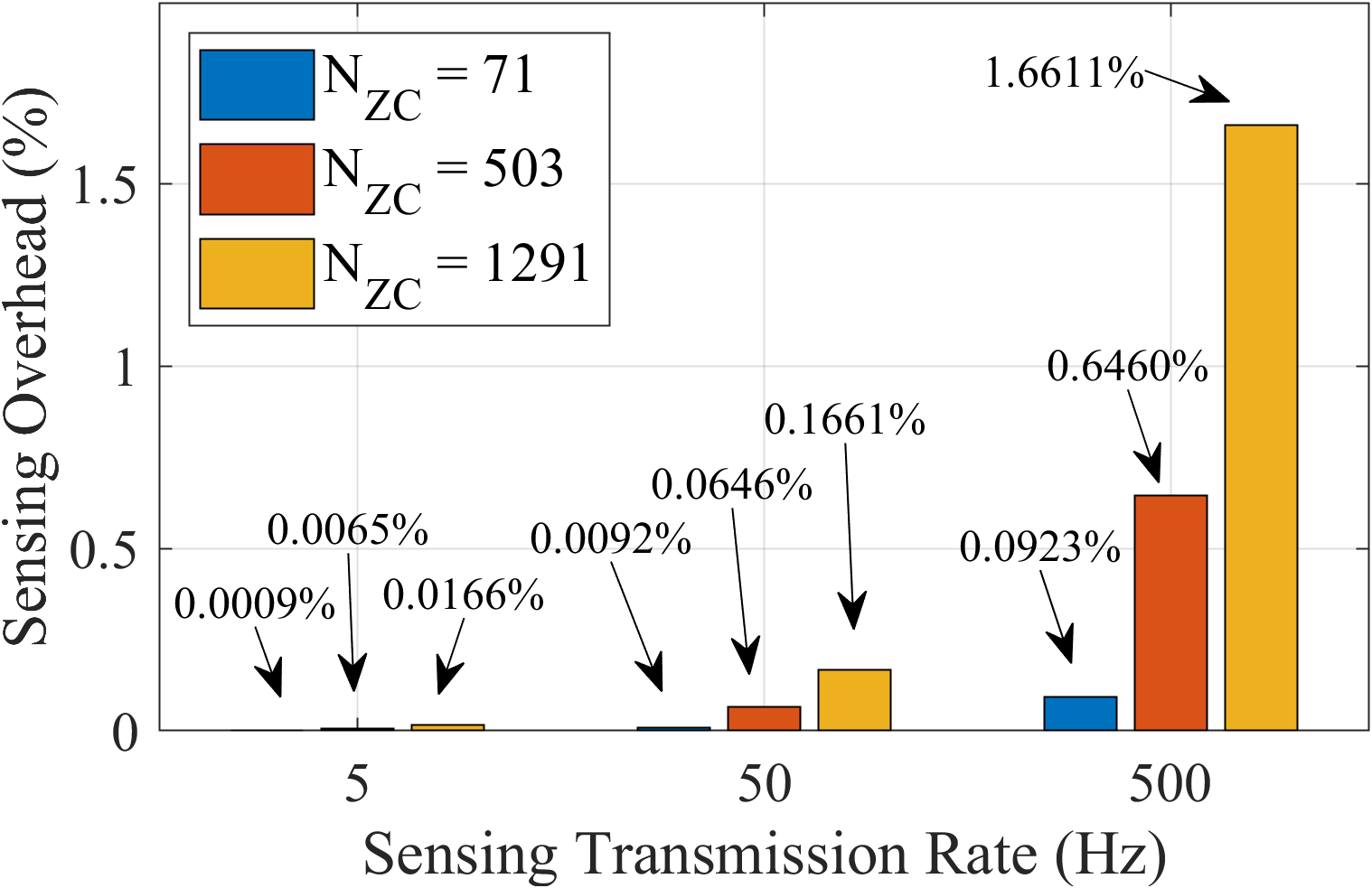}
    \caption{Single-beam sensing overhead per second (\% of shared resources remaining after fixed NR overhead) versus sensing transmission rate for representative ZC waveform lengths.}
    \label{fig:sensing_overhead}
\end{figure}

Fig.~\ref{fig:sensing_overhead} illustrates the sensing-resource overhead under the assumed 5G NR configuration. For a single sensing beam, the overhead remains modest across the considered waveform lengths and sensing rates. Even the most aggressive single-beam case ($N_\text{ZC}=1291$, $f_\text{sense}=500$ Hz) consumes only 1.66\% of the post-overhead shared resource budget $(1-\mathrm{OH})$, indicating that active sensing for a small number of targets can be supported with limited impact on communication throughput. The observed linear growth with beam count follows the conservative independent beam model in (\ref{eq:multibeam}); increasing the number of sensing beams from $N_b=1$ to $N_b=8$ while maintaining $N_\text{ZC}=1291$ and $f_\text{sense}=500$ Hz increases the sensing overhead to 13.29\%, leaving 86.71\% of schedulable resources available for communication. These results motivate load-aware adaptation of waveform length, sensing rate, and beam allocation rather than static sensing policies.

\subsection{Detection Performance with Adaptive Waveforms}

\begin{figure}[!t]
    \centering
    \includegraphics[width=\columnwidth]{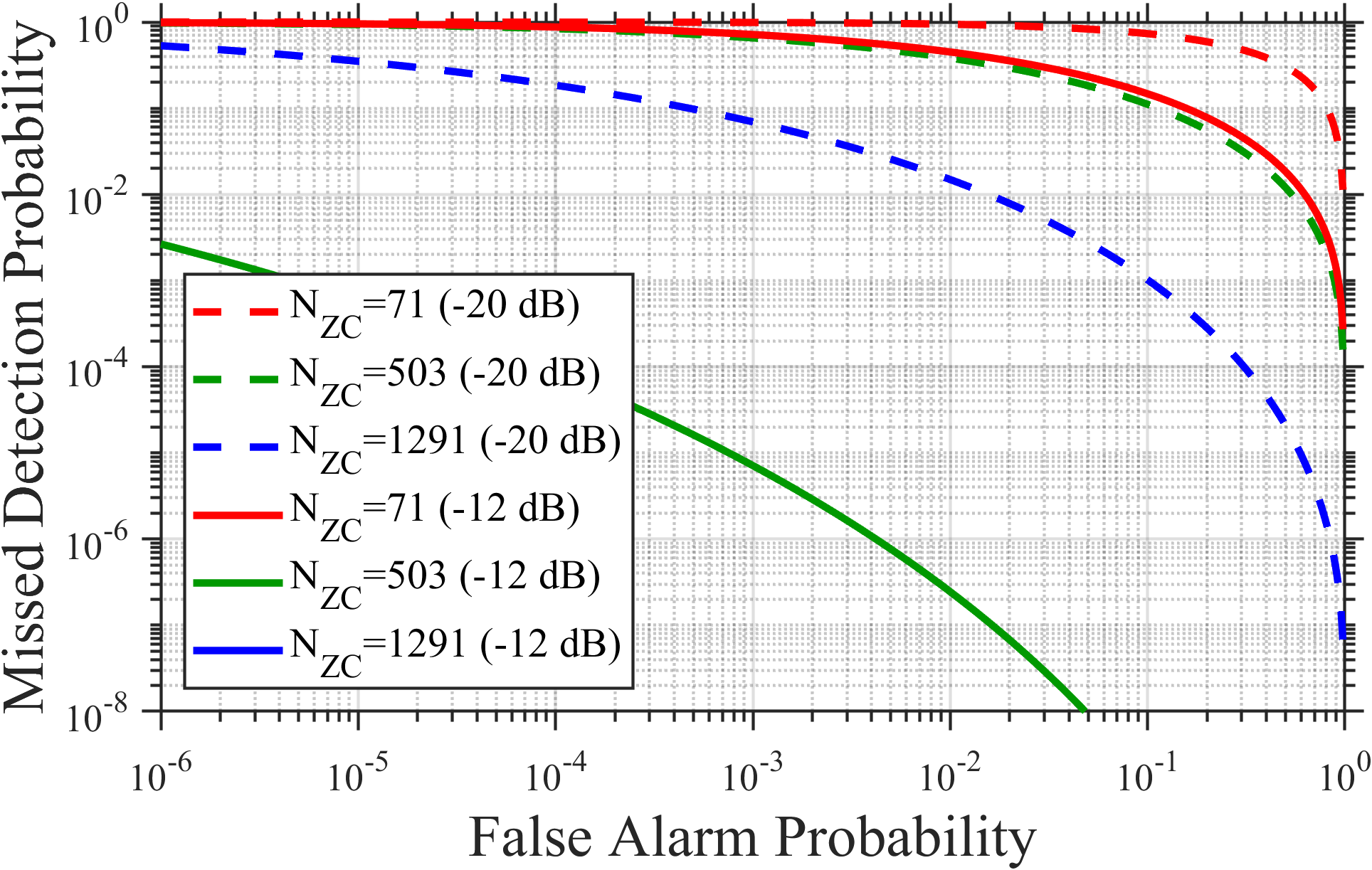}
    \caption{Missed-detection probability versus false-alarm probability for representative ZC waveform lengths under varying input SNR conditions.}
    \label{fig:roc_zc}
\end{figure}

Fig.~\ref{fig:roc_zc} presents receiver operating characteristic (ROC)-style performance through missed-detection probability versus false-alarm probability for representative ZC lengths under moderate SNR (-12 dB) and low SNR (-20 dB) sensing conditions. Consistent with the processing-gain model in Section~\ref{sensingmodel}, increasing $N_\text{ZC}$ lowers $P_\text{md}$ for a fixed $P_\text{fa}$ by increasing coherent processing gain, while higher input SNR improves detection performance across all waveform lengths. The benefit of increasing $N_\text{ZC}$ depends on the input SNR: under favorable conditions, shorter waveforms may provide sufficient detection reliability and preserve shared resources, whereas under degraded conditions, larger waveforms may be required to maintain reliable sensing performance. However, these gains diminish as SNR becomes severely degraded, where even the largest candidate waveform may provide limited improvement. Complementary sensing modalities such as SoO-assisted observations or multi-sensor fusion become increasingly valuable in severely degraded environments.

\subsection{Tracking Reliability Under Challenging SNR}

\begin{figure}[!t]
    \centering
    \includegraphics[width=\columnwidth]{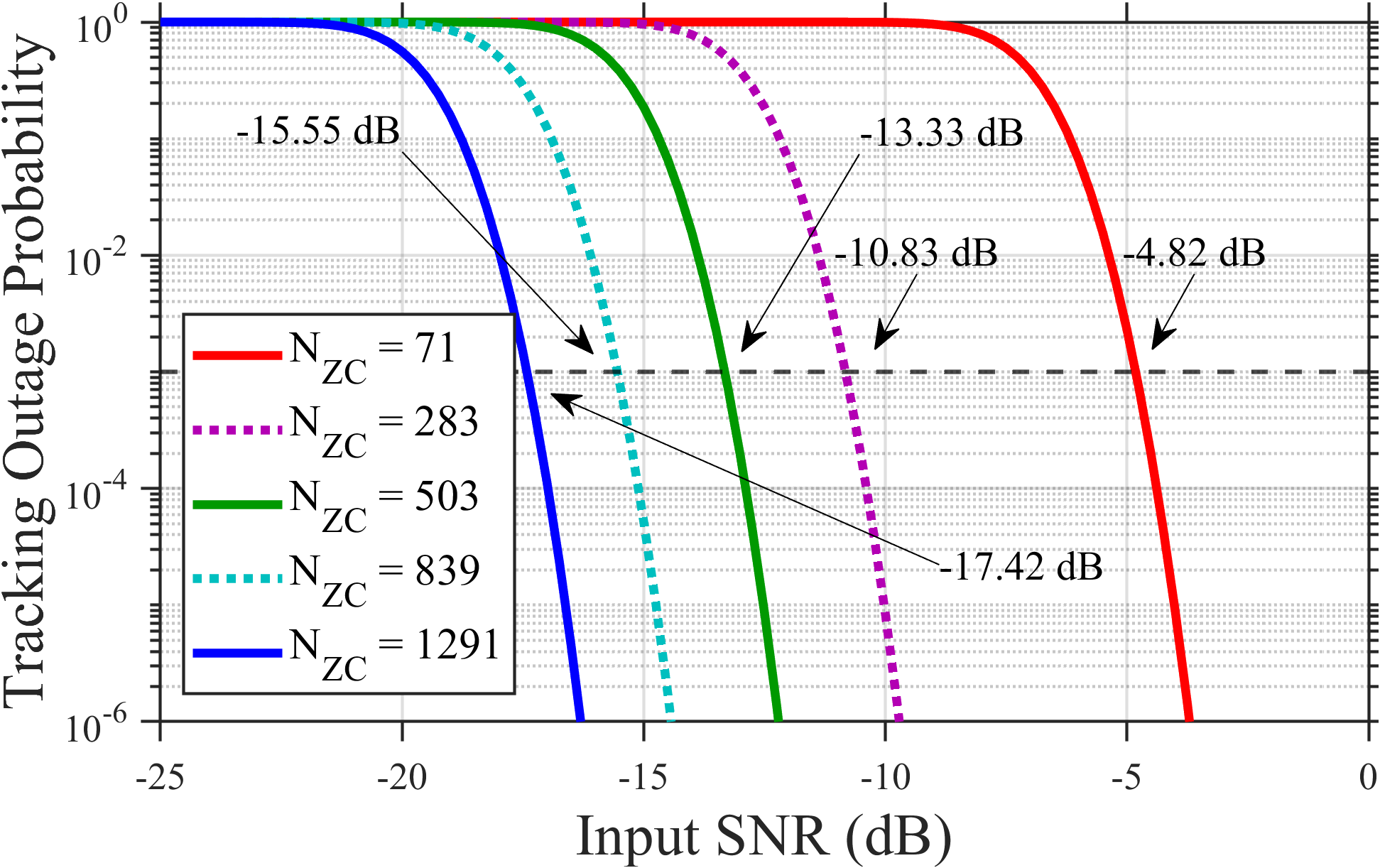}
    \caption{Tracking outage probability versus input SNR for representative ZC waveform lengths under the adopted $M=3$-of-$N=5$ tracking criterion with $P_{fa}=10^{-6}$.}
    \label{fig:tracking_outage_snr}
\end{figure}

Fig.~\ref{fig:tracking_outage_snr} shows tracking outage probability versus input SNR for representative ZC waveform lengths under the adopted $M$=3-of-$N$=5 tracking criterion with a constant $P_\text{fa}=10^{-6}$. Unlike single-opportunity detection metrics, tracking outage captures continuity over multiple sensing opportunities and exhibits a rapid decline in outage probability as input SNR increases. Consistent with the processing-gain model in Section~\ref{sensingmodel}, increasing $N_\text{ZC}$ shifts the outage curves left, indicating that the same tracking reliability can be achieved at lower SNR through coherent processing gain. For example, at a target outage probability of $10^{-3}$, the required input SNR improves from approximately -4.8 dB for $N_\text{ZC}=71$ to -17.4 dB for $N_\text{ZC}=1291$, which is a gain of roughly 12.6 dB. These results show that adaptive waveform selection can substantially extend reliable tracking into lower-SNR conditions, albeit at the cost of greater sensing-resource consumption.

\subsection{Load-Aware Throughput and Tracking Tradeoffs}

\begin{figure*}[!t]
\centering

\subfloat[Normalized throughput versus communication load.]{
    \includegraphics[width=0.48\textwidth]{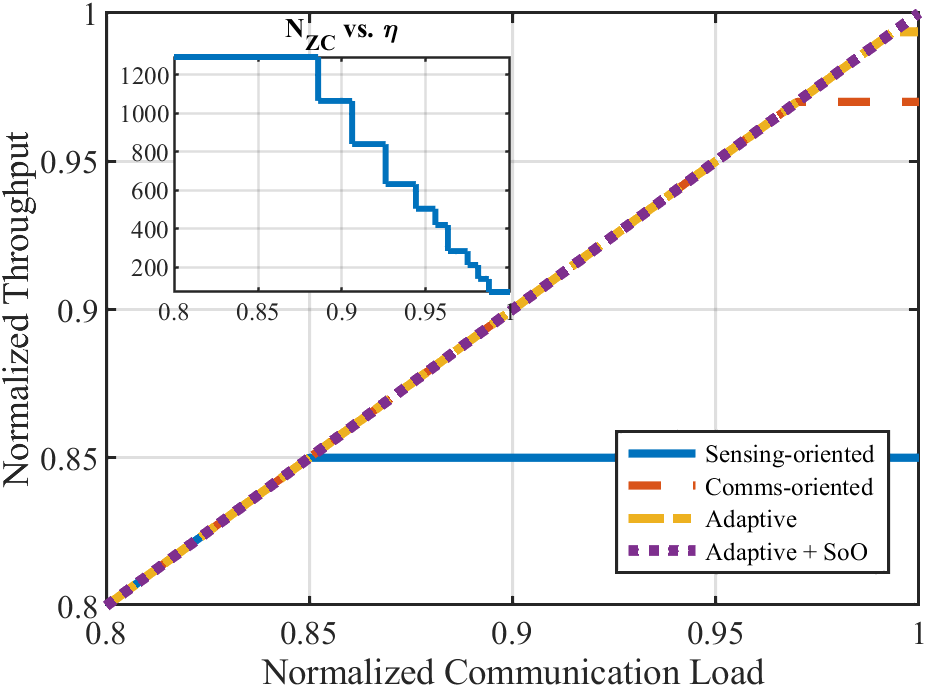}
    \label{fig:load_throughput}
}
\hfill
\subfloat[Tracking outage probability versus communication load.]{
    \includegraphics[width=0.48\textwidth]{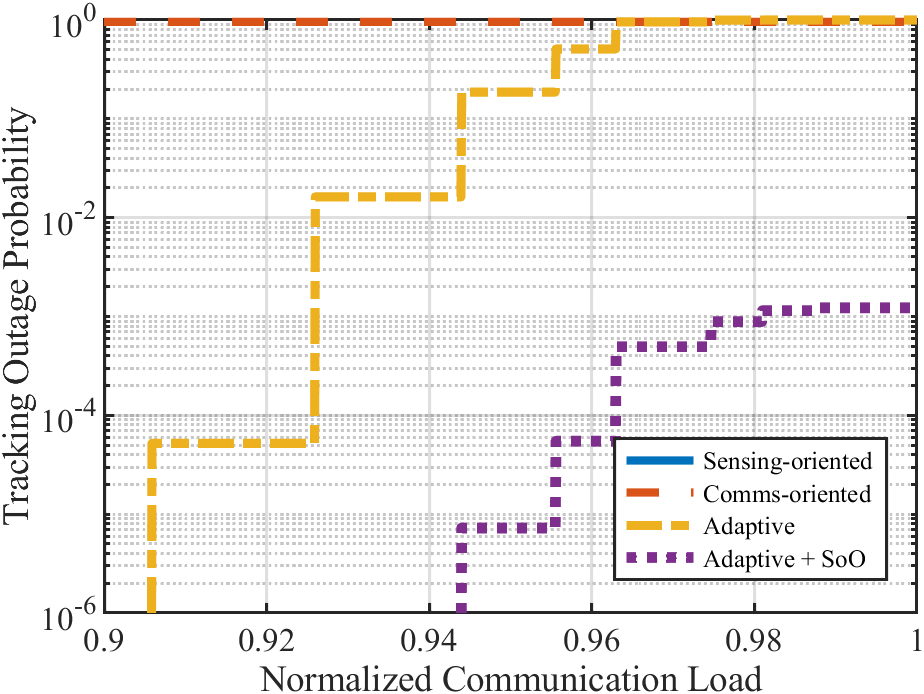}
    \label{fig:load_outage}
}

\caption{Throughput and tracking performance under heavy communication load for four resource-allocation policies: sensing-oriented, communications-oriented, adaptive 5G ISAC, and adaptive 5G ISAC with SoO assistance.}
\label{fig:load_tradeoff}

\end{figure*}

We next evaluate the four resource-allocation policies introduced in Section~\ref{commsmodel} under heavy communication load, with normalized offered traffic swept over $\eta \in [0.8, 1.0].$ An extreme sensing case is considered with $N_b=8$, corresponding to tracking multiple UAVs with eight sensing beams and $f_\text{sense}= 500 \text{ Hz}.$ Two fixed baselines are considered: a sensing-oriented policy that permanently reserves 15\% of the post-overhead shared budget for sensing, limiting normalized throughput to 0.85, and a communications-oriented policy that reserves 3\%, limiting throughput to 0.97. The adaptive policy dynamically adjusts sensing parameters to satisfy the current load, while the adaptive + SoO policy invokes distributed SDS nodes when active 5G sensing resources become constrained. Active 5G sensing uses $P_\text{fa}=10^{-6}$, input SNR = -15 dB, and an $M=3$ of $N=5$ tracking criterion.

Under the adaptive + SoO policy, SDS nodes exploit external SoO to provide supplemental passive sensing opportunities that augment active 5G ISAC sensing. Four SDS nodes are assumed, each contributing two additional sensing opportunities per observation window. Consequently, the number of passive sensing opportunities is $N_\text{SoO}=8$, yielding a total of $N_\text{tot}=N_{5\text{G}}+N_\text{SoO}=13$ sensing opportunities while retaining the requirement of $M=3$ successful detections for track continuity. Each SoO opportunity is modeled with an effective detection probability of $P_{\text{d,SoO}}=0.8$, representing reliable but non-ideal supplemental detections. Although different values of $P_{\mathrm{d,SoO}}$ would affect the magnitude of the observed gains, the qualitative benefits of SoO-assisted sensing remain unchanged. SoO assistance is assumed to be continuously available across the evaluated load range.

Fig.~\ref{fig:load_throughput} shows normalized throughput versus communication load for the four policies. Since throughput is normalized by abstracting spectral efficiency, the reported values represent the fraction of offered traffic supported by the remaining shared resource budget. The two fixed-reservation baselines saturate at their corresponding throughput limits of 0.85 and 0.97. The adaptive policy closely tracks the offered load over most of the operating range by progressively reducing $N_\text{ZC}$, as shown in the inset. However, any nonzero active sensing allocation introduces a throughput penalty under extreme communication loads. By offloading sensing from active 5G transmissions to passive SoO sensing when necessary, the adaptive + SoO policy maintains throughput equal to the offered load even under fully saturated conditions.

Fig.~\ref{fig:load_outage} illustrates the corresponding tracking outage probability versus communication load. The sensing-oriented policy maintains negligible outage because sufficient active sensing resources are always available, albeit at the cost of a 15\% reduction in communication throughput. In contrast, the communications-oriented policy preserves throughput but allocates insufficient resources to sensing, resulting in persistent tracking failure. The adaptive policy provides a favorable balance at moderate loads; however, tracking outage increases rapidly as active sensing resources are reduced to protect communication traffic. The adaptive + SoO policy mitigates this degradation by supplementing active 5G sensing opportunities with passive SoO observations, thereby maintaining low outage probabilities as the network approaches full utilization.

\subsection{Localization Accuracy of Multistatic Sensing}

Beyond throughput and tracking reliability, distributed sensing nodes can also improve estimation accuracy. Using the CRLB framework of Section~\ref{geometry}, PEBs are compared for monostatic and multistatic sensing geometries under different ZC sensing waveform lengths. Simulations consider a 2D horizontal slice at UAV altitude $z=100 \text{ m}$ over the region $[-500,500] \times [-500, 500] \text{ m},$ with the serving gNB located at (0,0,50) m and four SDS nodes located at $(\pm 300, \pm 300, 25)$ m. For clarity, the CRLB analysis assumes a single controlled illuminator (the serving gNB), while SDS nodes provide passive bistatic measurements only. External SoO transmitters are excluded to isolate sensing-geometry effects. The simulations assume $P_t =30$ dBm, $G_t = 20$ dBi, $\sigma_{\text{RCS}}=0.1$ m$^2$, $T = 290$ K, $F = 5$ dB, and $L = 0$ dB. Monostatic sensing assumes $G_{r}=20$ dBi and $\kappa_{\theta,m}=\kappa_{\phi,m}=0.07$, while bistatic SDS sensing assumes $G_{r}=10$ dBi and $\kappa_{\theta,b}=\kappa_{\phi,b}=0.28$ to model lower-cost distributed sensing nodes.

\begin{figure*}[!t]
    \centering
    \includegraphics[width=\textwidth]{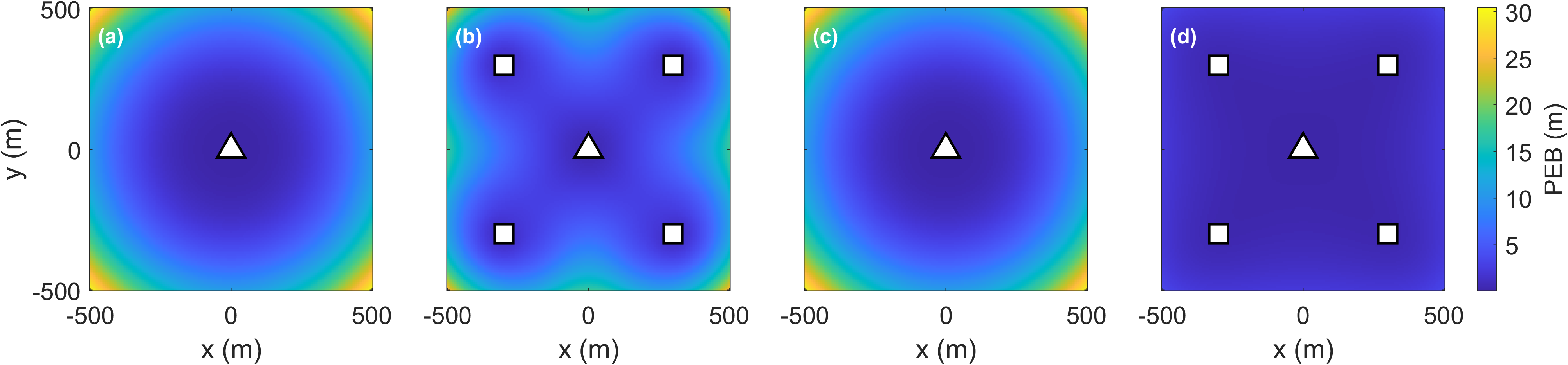}
    \caption{PEB comparison for monostatic and multistatic sensing geometries under different ZC sensing waveform lengths using the proposed CRLB framework. (a) Monostatic sensing with $N_{\text{ZC}}=71$. (b) Multistatic sensing with $N_{\text{ZC}}=71$. (c) Monostatic sensing with $N_{\text{ZC}}=1291$. (d) Multistatic sensing with $N_{\text{ZC}}=1291$. Lower PEB indicates better localization accuracy.}
    \label{fig:crlb_heatmaps}
\end{figure*}

Fig.~\ref{fig:crlb_heatmaps} compares PEB heatmaps for monostatic and multistatic sensing under short and long ZC sensing waveforms using the proposed SNR-aware CRLB framework. For the shorter waveform ($N_{\text{ZC}}=71$), monostatic and multistatic sensing provide comparable localization accuracy over much of the surveillance region, although the monostatic case degrades more rapidly toward the coverage edges due to the limited spatial diversity of a single observing node. Increasing the waveform length to $N_{\text{ZC}}=1291$ significantly improves multistatic localization accuracy by increasing the occupied sensing bandwidth and coherent processing gain across all bistatic sensing links. In contrast, the monostatic PEB improves only modestly, indicating that localization performance becomes increasingly geometry- and angle-limited for a single observing node. Averaged over the surveillance region, increasing the ZC sequence length from $N_{\rm ZC}=71$ to $N_{\rm ZC}=1291$ reduces the mean multistatic PEB from 6.19 m to 0.85 m, while the monostatic mean PEB remains approximately constant at 6.95 m and 6.88 m, respectively. These results highlight how distributed multistatic sensing geometries more effectively exploit improvements in sensing bandwidth and waveform processing gain than monostatic sensing alone. 

\section{Conclusion}

This paper investigated adaptive multistatic ISAC for load-aware UAV detection and tracking in congested RF environments. A shared-resource framework quantified how waveform selection, sensing rate, and beam allocation affect communication performance in 5G NR systems. Results showed that adaptive sensing outperforms fixed reservations by preserving throughput while maintaining sensing capability. Under severe congestion, distributed SDS nodes leveraging SoO further reduced tracking outage. CRLB analysis also showed that multistatic sensing improves localization accuracy and spatial uniformity relative to monostatic sensing. Future work will incorporate antenna patterns, directional beamforming, and propagation-dependent channel effects into the models and validate the framework on the NSF AERPAW testbed.



\bibliographystyle{IEEEtran}
\bibliography{references.bib}

\end{document}